\documentclass[10pt,conference]{IEEEtran}

\usepackage{amsmath,amssymb}
\usepackage{graphicx}
\usepackage{cite}
\usepackage[latin1]{inputenc}
\usepackage{color}
\usepackage{multirow}
\usepackage{siunitx}

\definecolor{light-gray}{rgb}{0.8,0.8,0.8}
\definecolor{BrickRed}{rgb}{0.8,0.0,0.0}
\definecolor{Brown}{rgb}{0.4,0.2,0.2}

\IEEEoverridecommandlockouts
\newcommand{\norm}[1]{\left\lVert{#1}\right\rVert}

\newcommand{\argmax}[1]{\underset{#1}{\operatorname{argmax} \ }}

\def\bsalpha{{\boldsymbol{\alpha}}}

\def\bszeta{{\boldsymbol{\zeta}}}

\def\bssigma{{\boldsymbol{\sigma}}}

\def\bsp{{\mathbf{p}}}

\def\bsG{{\mathbf{G}}}

\def\nr{{N_\text{r}}}

\def\nm{{N_\text{m}}}

\begin{document}
\title{Participatory Sensing for Localization of a GNSS Jammer}

\author{
	\IEEEauthorblockN{ Gl\"adje Karl Olsson\IEEEauthorrefmark{1}, Erik 
Axell\IEEEauthorrefmark{1}, Erik G. 
Larsson\IEEEauthorrefmark{2} and Panos
		Papadimitratos\IEEEauthorrefmark{3}}
	\IEEEauthorblockA{\IEEEauthorrefmark{1}Swedish Defence Research Agency 
	(FOI),
		Link\"oping, Sweden\\ 
		{\{%
		gladje.karl.olsson, erik.axell%
		\}}{@foi.se}
	} \IEEEauthorblockA{\IEEEauthorrefmark{2}Link\"oping University,
	Link\"oping, Sweden\\
	{
	erik.g.larsson%
}{@liu.se}
	} \IEEEauthorblockA{\IEEEauthorrefmark{3}Royal Institute of Technology 
(KTH),
Stockholm, Sweden\\
{
	papadim%
}{@kth.se}
}
\thanks{This work was supported in part by Security Link}
}

\IEEEoverridecommandlockouts

\IEEEpubid{\begin{minipage}{\textwidth}
\copyright2022 IEEE. Personal use of this material is permitted. Permission 
from \\
IEEE must be obtained for all other uses, in any current or future media, \\
including reprinting/republishing this material for advertising or promotional 
\\
purposes, creating new collective works, for resale or redistribution to 
servers \\
or lists, or reuse of any copyrighted component of this work in other works.
\end{minipage}}

\maketitle

\IEEEpubidadjcol

\begin{abstract}
GNSS receivers are vulnerable to jamming and spoofing attacks, and numerous 
such incidents have been reported worldwide in the last decade. It is important
to detect attacks fast and localize attackers, which can be hard if not impossible 
without dedicated sensing infrastructure.
The notion of participatory sensing, or crowdsensing, is that a large ensemble
of voluntary contributors provides the measurements, rather than relying on
dedicated sensing infrastructure. This work considers embedded GNSS receivers 
to provide measurements for participatory jamming detection and localization.
Specifically, this work proposes a novel jamming localization algorithm, based on
participatory sensing, that exploits AGC and $C/N_0$ estimates from commercial
GNSS receivers. The proposed algorithm does not require knowledge of the jamming
power nor of the channels, but automatically estimates all parameters.
The algorithm is shown to outperform similar state-of-the-art localization
algorithms in relevant scenarios.
\end{abstract}

\begin{IEEEkeywords}
GNSS, jamming, participatory sensing, crowdsensing
\end{IEEEkeywords}

\section{Introduction}
\label{sec:introduction}

Global Navigation Satellite System (GNSS) receivers are widely spread in 
numerous society-critical services today. At the same time, GNSS receivers are 
vulnerable to jamming and spoofing attacks, and numerous GNSS jamming and 
spoofing incidents have been reported worldwide in the last decade. Detecting, 
and more importantly, localizing the source of such attacks is therefore of 
significant importance.

Considering that any GNSS equipped device could be targeted by an attack, 
measurements available at GNSS receivers can be the basis for the 
detection and localization of an attack at each device. Typically, 
GNSS receivers are embedded in a wide gamut of networked devices (e.g., 
smart-phones, vehicles). This enables each affected device to share data 
of such an event with other devices, typically with the help of a data 
aggregating service. This is the notion of participatory sensing, or 
crowdsensing: rather than relying on dedicated sensing infrastructure, 
a large ensemble of voluntary contributors provides the necessary 
measurements. In our scenario, the sensors are GNSS receivers embedded in a 
conncected device and measurements refer to the data available in such GNSS 
receivers. GNSS receivers offer a rich interface to radio measurements that can 
be valuable in detecting attacks and localizing attacking devices. This 
motivated a number of recent publications that deal with detection or 
localization of GNSS interference using a crowdsensing approach (cf. 
\cite{Borio-GNSS+-2016, Ahmed-ICL-GNSS-2020, 
Ahmed-AeroConf-2021, Liu-CSNC-2020, Wang-TAES-2020}). 
These works are based on some sort of power measurement, usually through 
carrier-to-noise-density ratio ($C/N_0$) estimates or automatic gain 
control (AGC) values, but in some cases through direct power measurements. 
$C/N_0$ measurements, and sometimes AGC, are provided by all grades of 
GNSS receivers, from low-cost to professional, which allows jammer 
localization using different types of commercial sensors.

\cite{Borio-GNSS+-2016, Ahmed-ICL-GNSS-2020, Ahmed-AeroConf-2021, Liu-CSNC-2020}
utilizes the $C/N_0$ and a simple distance dependent path loss model, and
computes the jammer position based on a least squares (LS) solution for multiple
receivers. The LS localization algorithm of \cite{Borio-GNSS+-2016} is derived 
for
a single moving receiver which can be viewed as a synthetic array. An algorithm
for multiple receivers, that takes the average location of
those receivers that detect the jammer, is also proposed in 
\cite{Borio-GNSS+-2016}. This algorithm is further extended in
\cite{Wang-TAES-2020} to obtain a weighted average solution, rather than just a
plain average. 

There are other related papers that deal with power difference of
arrival (PDOA) algorithms \cite{Tucker-GNSS+-2020}. These are similar to
exploiting AGC or $C/N_0$ measurements, but they are assumed to measure, or
estimate, the received power directly. Some of these even assume that the power
difference is estimated from the AGC \cite{Tucker-GNSS+-2020, Blay-ITM-2018}.

The model assumptions of \cite{Borio-GNSS+-2016, Ahmed-ICL-GNSS-2020, 
Ahmed-AeroConf-2021, Wang-TAES-2020, Liu-CSNC-2020, Tucker-GNSS+-2020, 
Blay-ITM-2018} are very similar to those used in this
work. The main differences are that our work extends the models of 
\cite{Borio-GNSS+-2016, Ahmed-ICL-GNSS-2020, Ahmed-AeroConf-2021, 
Wang-TAES-2020, Liu-CSNC-2020, Tucker-GNSS+-2020, Blay-ITM-2018} by assuming an 
unknown channel (i.e., the path
loss exponent) between the receivers and jammer and a random
measurement error is included. This is necessary for practical application. The 
work of \cite{Blay-ITM-2018} also deals with 
an unknown path loss exponent using time difference of arrival (TDOA). That 
approach, however, requires access to baseband I/Q data which 
are not available in a standard embedded GNSS receiver and is therefore not a 
suitable algorithm for a participatory sensing scenario. In addition, our
proposed algorithm does not require pre-calibration of the sensing receivers
with respect to the jamming power, contrary to the algorithms of
\cite{Borio-GNSS+-2016, Ahmed-ICL-GNSS-2020, Ahmed-AeroConf-2021,
	Wang-TAES-2020}.

Mobile phone based crowdsourcing for jamming detection and localization is
considered in \cite{Strizic-ITM-2018, Kraemer-GNSS-2012}. The basic concept of
using AGC or $C/N_0$ measurements from mobile phones for localization purposes
is proposed in \cite{Strizic-ITM-2018}, and field trials show the usefulness of
these metrics. However, no explicit localization algorithm is proposed in 
\cite{Strizic-ITM-2018}. \cite{Kraemer-GNSS-2012} proposes a localization 
algorithm based on $C/N_0$ measurements in combination with step detection 
and step length estimation using an inertial sensor. The algorithm of 
\cite{Kraemer-GNSS-2012} assumes 
a sensor that moves along a straight line in two perpendicular directions, 
which is highly limiting in a participatory sensing system.

The main contribution of this work is a novel jamming 
localization algorithm, based on participatory sensing, that
\begin{itemize}
	\item does not need any pre-calibration or channel knowledge but 
	automatically estimates all parameters,
	\item exploits AGC or $C/N_0$ estimates, or a combination thereof, from 
	commercial GNSS receivers,
	\item outperforms existing similar algorithms in relevant scenarios.
\end{itemize}
\section{System Model}
\label{sec:model}
 
We do not dwell in this paper on the networking specifics or the security and 
privacy of the collected data (e.g., \cite{GisdakisGP:J:2016, GisdakisGP:C:2015}). 
Rather, we assume $\nr$ receivers measure AGC and $C/N_0$ and submit those to a 
central server, where the localization algorithm is run. Measurements are 
submitted at a specified common rate, typically in the order of 1 Hz from a 
standard mobile phone. Submitted data are assumed to be either time-stamped 
or can be otherwise synchronized at an accuracy within the common time epoch 
of each measurement.

The proposed algorithm is based on the following assumptions:
\begin{itemize}

\item the receivers have isotropic antennas,
\item the jammer has an isotropic antenna,
\item the jammer position, denoted by $\bsp_0 = [x_0, y_0, z_0]^T$, is fixed 
during the time of mesurement,
\item the jamming power, denoted by $J_0$, is constant during the time of 
mesurement,
\item receiver noise powers are constant during the time of 
mesurement,
\item the receiver positions, denoted by $\bsp_i[n] = [x_i[n], y_i[n], 
z_i[n]]^T$ at time $n$ for receiver $i=1, \ldots,\nr$, are known.
\end{itemize}
These assumptions are also made (explicitly or implicitly) in
\cite{Borio-GNSS+-2016, Ahmed-ICL-GNSS-2020, Ahmed-AeroConf-2021,
Wang-TAES-2020}. It should be noted that
the assumptions should be valid only within each measurement time frame, which
is a design parameter. The assumption of isotropic antennas boils down to an
assumption of the channel gains being constant during the time of measurement,
which is required for estimating the channel. Constant noise powers within each
measurement time frame is a fair approximation for the relatively short
measurement times that are of interest. Constant jamming power is valid for most
commercial jammers. The receiver positions can be assumed known by integration of
GNSS and other sensors, such as an inertial measurement unit (IMU),
embedded in the sensor device (e.g. a cell phone). That is, at any point in 
time the sensing device has a current own position estimate it uses to 
geostamp the contributed measurement. It should be noted that the
proposed algorithm estimates a jammer position even if these assumptions do not
hold perfectly, but of course the estimation error becomes larger if, for
example, the jammer moves during the observation time.

The goal is to estimate 
the
position $\bsp_0$ of the jammer.
The model for AGC measurements is explained first, followed by an analogous 
model for $C/N_0$ estimates.

\subsection{AGC}
Let $h_i[n]$ denote 
the channel (power) gain between the attacker and receiver $i$ at time $n$. 
The channel gain $h_i[n]$ depends on the distance 
$d_i[n] \triangleq \norm{\bsp_0 - \bsp_i[n]} = \sqrt{(x_0-x_i[n])^2 + 
(y_0-y_i[n])^2 + (z_0-z_i[n])^2}$). 
Then, the received jamming power $J_i[n]$ is
\begin{equation*}
	J_i[n] = J_0 h_i[n].
\end{equation*}
A path loss model where $h_i[n] \propto 
1/d_i[n]^{\alpha_i}$ yields
\begin{equation*}
	J_i[n] = \frac{J_0 \kappa}{d_i[n]^{\alpha_i}},
\end{equation*}
where $\alpha_i$ is the path loss exponent for 
receiver $i$ and $\kappa$ is a 
proportionality
constant. The constant $\kappa$ depends on, for example, carrier frequency and 
antenna gain, which are constant and equal for all receivers as all antennas 
are assumed to be isotropic. The path loss exponents $\alpha_i$ are assumed to 
be unknown in this work and are estimated as explained in 
Section~\ref{sec:estimation}.

The total received power, denoted by $P_i[n]$, can 
be written
\begin{equation*}
	P_i[n] = J_i[n] + N_{G,i} + \sum_{j=1}^{\nm}C_{i,j}[n],
\end{equation*}
where $N_{G,i}$ is the background noise power at the 
input to the AGC circuit 
of receiver $i$, 
assumed to be constant during the time of 
measurement, and $C_{i,j}[n]$ denotes 
the power of the
visible satellite signal $j=1, \ldots, \nm$ at receiver $i$. The received 
powers of the 
satellite signals (before despreading)
are small compared to the receiver noise power and can therefore be 
neglected, i.e.,
\begin{equation*}
	P_i[n] \approx J_i[n] + N_{G,i} = \frac{J_0 
	\kappa}{d_i[n]^{\alpha_i}} + N_{G,i}.
\end{equation*}

Let $g_i[n]$ be the AGC value for receiver $i = 1, 
\ldots, \nr$ at time $n$. 
Then, ideally, 
(\cite{Thompson-IGNSS-2013, Thompson-PhD-2013})
\begin{equation*}
	g_i[n] \sqrt{P_i[n]} = \sqrt{\bar P_i},
\end{equation*}
where $\sqrt{\bar P_i}$ is the desired received signal amplitude. 
The AGC gain in the absence of jamming ($J_0=0$), denoted by $\bar g_i$, is assumed to be
known and can be written
\begin{equation*}
	\bar g_i\sqrt{N_{G,i}} = \sqrt{\bar P_i}.
\end{equation*}
In practice, $\bar g_i$ can easily be estimated at initialization or by 
long-term estimation.
It should be noted that the algorithm uses the difference relative to
the non-jammed AGC value, $\bar g_i$, and not the absolute measurement.
This is important, because the absolute AGC estimates may vary
significantly across devices \cite{Lee-ITM-2021}.
Then,
\begin{equation*}
	g_i[n] = \frac{\sqrt{\bar P_i}}{\sqrt{P_i[n]}} = \bar 
	g_i\sqrt{\frac{N_{G,i}}{P_i[n]}} 
=\bar g_i 
\sqrt{\frac{N_{G,i}}{\frac{J_0 \kappa}{d_i[n]^{\alpha_i}} + N_{G,i}}}.
\end{equation*}

Let $G_i[n]$ and $\bar G_i$ denote $g_i[n]$ and $\bar 
g_i$, respectively, in decibels. That is, $G_i[n] \triangleq 
20\mathrm{log}_{10}(g_i[n])$ and 
$\bar G_i \triangleq 20\mathrm{log}_{10}(\bar g_i)$. Then, 
the measurement 
model, with additional measurement noise, can be 
written as
\begin{equation}
	\label{eq:AGC}
G_i[n] = \bar G_i - 
10\mathrm{log}_{10}\left(\frac{J_0\kappa}{N_{G,i}}d_i[n]^{-\alpha_i} + 1\right) 
+ w_{G_i}[n],
\end{equation} 
where $w_{G_i}[n] \sim \mathcal{N}(0,\,\sigma_{G,i}^{2})$ 
denotes measurement 
noise.

\subsection{$C/N_0$}
Let $s_{i,j}[n]$ denote the observed 
carrier-to-noise-and-interference
ratio (CNIR), which is estimated by the GNSS receivers and commonly, with slight abuse of terminology, referred to as 
$C/N_0$. Let $N_{0,i}$ denote the background noise power spectral 
density at receiver $i$, and $\tilde J_i[n]$ a spectrally-flat-equivalent 
interference noise power density \cite{Murrian-GNSS+-2019} at receiver $i$. 
Then, the CNIR can be written as
\begin{equation*}
s_{i,j}[n] \triangleq \frac{C_{i,j}[n]}{N_{0,i} + \tilde J_i[n]} = 
\frac{C_{i,j}[n]}{N_{0,i}} \cdot \frac{1}{1 + \frac{\tilde 
J_i[n]}{N_{0,i}}}.
\end{equation*}
Multiplying $\tilde J_i[n]$ and 
$N_{0,i}$ in the denominator with the receiver bandwidth, thereby cancelling 
the bandwidth dependence, and taking the mean CNIR values from all received 
satellite signals yields
\begin{equation*}
s_{i}[n] = \frac{1}{N_m}\sum_{j=1}^{N_m}\left(\frac{C_{i,j}[n]}{N_{0,i}} \cdot \frac{1}
{1 + \frac{J_i[n]}{N_{S,i}}}\right),
\end{equation*}
where $N_{S,i}$ is the noise power experienced in CNIR estimation. Note that the
noise power at the CNIR estimator, $N_{S,i}$, is not in general equal to the AGC 
input noise, $N_{G,i}$. $N_{S,i}$ includes additional noise from circuits (such as
low-noise amplifiers) inbetween the AGC and the CNIR estimator. Moreover, it is 
estimated in the post-correlation step, whereas the $N_{G,i}$ noise is experienced 
on the pre-correlation input signal.

Let $\bar s_{i}$ be the mean CNIR with no jamming ($J_i[n] = 0$), which is 
assumed to be known. In practice, $\bar s_{i}$ can easily be estimated at 
startup (assuming there is no jamming) or by long-term estimation 
of the measured CNIR. That is
\begin{equation*}
	\bar s_{i} = \frac{1}{N_m}\sum_{j=1}^{N_m}\frac{C_{i,j}[n]}{N_{0,i}}.
\end{equation*}

Recall that the jammer localization algorithm uses the difference
relative to the non-jammed value $\bar s_{i}$ and not the measurement itself in
absolute numbers, to avoid significant variations across devices.
Then, 
\begin{equation*}
	s_{i}[n] = \bar s_{i} \cdot  \frac{1}{1 + 
	\frac{J_i[n]}{N_{S,i}}}.
	\end{equation*}

By switching to logarithmic scale, i.e. using $S_{i}[n]$ and $\bar S_{i}$ to
denote $s_{i}$ and $\bar s_{i}$, respectively, in decibels, the 
measurement model is written as
\begin{equation}
\label{eq:cn0}
S_{i}[n] = \bar S_{i} - 10\mathrm{log}_{10}\left(\frac{J_0\kappa}{N_{S,i}}d_i^{-\alpha_i}[n] + 1\right) + 
w_{S_{i}}[n],
\end{equation} 
where $w_{S_{i}}[n] \sim \mathcal{N}(0,\,\sigma_{S_{i}}^{2})\,$, which is 
analogous to \eqref{eq:AGC}. From here on, only the AGC measurement 
model will be used, but the CNIR measurement model can be used in the same way 
since equations~\eqref{eq:AGC} and \eqref{eq:cn0} are identical.

\section{Jammer Position Estimation}
\label{sec:estimation}
We derive an algorithm that estimates the unknown jammer position $\bsp_0$. In addition 
to $\bsp_0$, there are several unknown nuisance parameters that need 
to be taken into account. To simplify notation, the model is parametrized as 
follows:
\begin{align*}
	&\zeta_i \triangleq \frac{J_0\kappa}{N_{G,i}}, \ \ \bszeta \triangleq 
	[\zeta_1, \ldots, \zeta_\nr]^T, \\
	&\bsalpha \triangleq [\alpha_1, \ldots, \alpha_\nr], \ \
	\bssigma_{G}^{2} \triangleq [\sigma_{G,1}^{2}, \ldots, 
	\sigma_{G,\nr}^{2}]. \\
\end{align*}

\subsection{Joint Estimation} 
To perform joint estimation 
of the position based on all jammed sensors, $\nr$, the AGC observations for 
time instances $n=1,\ldots, N$ are combined in the matrix
\begin{equation*}
	\bsG \triangleq
	\begin{bmatrix}
		G_1[1] & G_2[1] & \ldots & G_\nr[1] \\
		G_1[2] & G_2[2] & \ldots & G_\nr[2] \\
		\vdots    & \vdots    & \ddots & \vdots      \\
		G_1[N] & G_2[N] & \ldots & G_\nr[N] \\		 
	\end{bmatrix},
\end{equation*}
and the known AGC levels in the absence of jamming in the vector
\begin{equation*}
	\bar \bsG \triangleq
	\begin{bmatrix}
		\bar G_1 & \bar G_2 & \ldots & \bar G_\nr	 
	\end{bmatrix}.
\end{equation*}

The likelihood function for the AGC measurements is then
\begin{equation}
	\label{eq:agc_likelihood}
	\begin{split}
		&p(\bsG|\bar \bsG, \bsp_0, \bsalpha, \bszeta, \bssigma_{G}^{2}) = \\
		&= \prod_{i=1}^\nr (2\pi\sigma_{G,i}^{2})^{-N/2}\mathrm{exp} \Big[ -\frac{1}{2\sigma_{G,i}^{2}} \sum_{n=1}^N ( 
		G_i[n] -  \\
		& (\bar G_i - 10\mathrm{log}_{10}(\zeta_i d_i^{-\alpha_i}[n] + 1)) )^2 \Big].
		\end{split}
\end{equation}

The jammer position is found by maximizing the likelihood function as
\begin{equation}
	\label{eq:max_agc}
	\bsp_0^* = \argmax{\bsp_0, \bszeta, \bssigma_{G}^{2},\bsalpha} p(\bsG|\bar \bsG, \bsp_0, 
	\bsalpha, \bszeta, \bssigma_{G}^{2}).
\end{equation}

\subsection{Estimation from a Subset of Sensors}

The joint estimation may not perform well if some of the sensors are 
 unreliable or do not fit the model well. This occurs, for example, if the 
channel model $h_i[n] \propto 1/ d_i[n]^{\alpha_i}$ is inaccurate for some 
sensors, especially if the jamming signal is obstructed. Many other 
algorithms, such as \cite{Ahmed-ICL-GNSS-2020, Borio-GNSS+-2016}, assume that 
$\alpha_i = 2$ which is true in ideal free space propagation. In practical 
cases, the path loss may differ from 2 and hence cause, eventually, significant 
sensor-jammer distance estimation errors. Therefore, the final position 
estimate is based on the subset of the 
available sensors deemed to be more accurate, thus more useful. Receivers that 
have free sight to the jammer and hence $\alpha_i \approx 2$ are most useful 
in terms of accurately estimating the distance to the jammer. Therefore, in 
this work, receivers with estimated $\alpha_i \leq 2.3$ are deemed useful 
for the localization estiamation.

To select the desired set of most useful sensors, the path loss exponent 
$\alpha_i$ is estimated for each sensor individually by maximizing the 
likelihood function using that particular receiver only, and different values 
of $\alpha_i$ using a grid search. The value of $\alpha_i$ that results in 
the highest likelihood is then chosen as the path loss exponent estimate for 
that receiver. The joint position estimate is then calculated by maximizing 
the likelihood function, using only the most useful receivers with their 
estimated pathloss exponents.

\subsection{Maximizing the Likelihood Function}

The maximization \eqref{eq:max_agc} cannot be solved analytically, but 
must be solved using numerical methods. 
The likelihood function is maximized using gradient descent on the negative 
log-likelihood. The parameter $\bssigma_{G}$ can, however, 
be solved analytically, conditioned on the other unkown parameters, as
\begin{equation*}
	\sigma_{G,i}^2 = 
	\frac{1}{N}\sum_{n=1}^N(G_i[n]-\bar{G}_i+10\log_{10}(\zeta_id_i^{-\alpha_i}[n]+1))^2
\end{equation*}
when using AGC.

Hence, the $\bssigma_{G}$ is computed 
according to this expression after each iteration of the gradient descent.

The initial jammer position estimate is set to the mean start position of 
the receivers, i.e., $\bsp_0 =  \frac{1}{\nr}\sum_{i=1}^{\nr}\bsp_i[0]$, 
and $\bssigma_{G}^{2}$ is initiated to the variance of the AGC value of 
each receiver when no jamming is present. During the pathloss exponent 
estimation, $\bszeta$ is initialized to \SI{e8}{}. The resultant 
$\bszeta$ values after the path loss estimation gradient descent are used 
as initial values during the joint estimation.

In order to find a suitable stepsize during the descent, the backtracking line 
search is utilized, first proposed in \cite{Armijo-pjm-1966}. The path loss exponent, $\bsalpha$, is optimized through grid 
search and is not a part of the gradient descent. That is, $\bsalpha$ is chosen 
beforehand from a grid and then held fixed during the gradient descent 
computation. The reason for this is that the gradient descent does not converge 
well to the correct value of $\bsalpha$ in general.

\section{Numerical Results}

The proposed algorithm is evaluated numerically, and compared to state-of-the-art methods. 
Monte Carlo simulations are used to evaluate the performance of the 
algorithms. For each Monte Carlo-run, the start positions of the 
receivers are placed at a uniformly random point within a cube with a side 
length of 2000 meters, with the jammer placed at the center of the cube.
The jammer transmits a signal with power \SI{0.01}{\watt} over the same 
bandwidth used by the victim receivers. 

The scenario starts at a random time of day and goes on for a predefined length 
of time. The receivers are mobile, travelling from their start point in a random 
direction with a preset constant speed and the jammer is activated halfway 
through the scenario. The samples from the first half of the scenario, where 
no jamming is present, is used to estimate $\bar G_i$. The samples from the 
second half of the scenario is used as the AGC observations, $\bsG$, in the
likelihood function \eqref{eq:max_agc}. If a measurement from a receiver drops 
down -5dB from $\bar G_i$ at any time during the scenario, the receiver is 
considered to be jammed and it is used in the joint estimation. The path loss 
exponent between each receiver and the 
jammer is set to 2, with an added random error taken from a half-normal 
distribution with scale parameter $0.1$.

In Sections \ref{sec:receivers} to \ref{sec:zeta}, the AGC model 
\eqref{eq:agc_likelihood} is used and compared to the algorithms presented in 
\cite{Ahmed-ICL-GNSS-2020, Borio-GNSS+-2016}, for
GNSS jamming localization based on crowdsourced measurements, 
that is, the state-of-the-art. Both algorithms are based on similar properties 
as the algorithm proposed in this work. The main difference between the two 
compared algorithms and this work is that the proposed novel algorithm also 
considers measurement noise and unknown path loss exponents, which is necessary 
for practical application.
To do a fair comparison, the compared algorithms are modified to use AGC measurements 
in lieu of CNIR. The algorithm of \cite{Borio-GNSS+-2016} estimates the jammer 
position as the mean of the positions of all detecting receivers. The algorithm 
of \cite{Ahmed-ICL-GNSS-2020} is based on a least-squares estimate. For the 
method in 
\cite{Ahmed-ICL-GNSS-2020}, the receivers are calibrated with a known 
jamming power observed at a known distance and the path loss exponents are 
assumed to be $\alpha_i = 2$, even when the actual path loss differs from $2$ 
in the simulations.

\subsection{Different Number of Receivers}\label{sec:receivers}
\begin{figure}[tbp]
	\includegraphics[width=1.0\columnwidth]{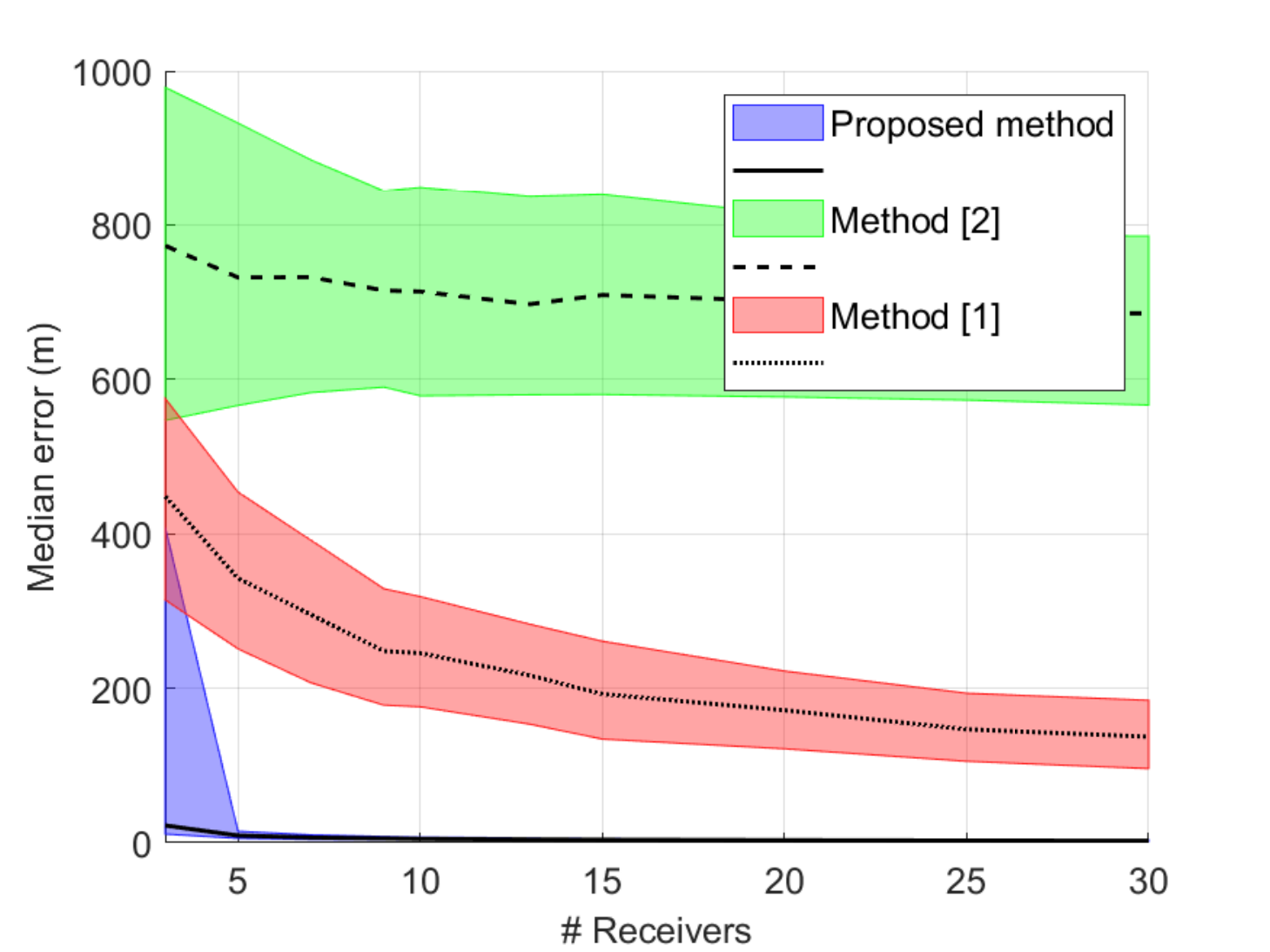}
	\caption{Median 3D-error (black lines) and 25th to 75th percentile (shaded 
	areas) for the three different methods using different number of receivers.}
	\label{fig:receivers}
\end{figure}

\begin{figure}[tbp]
	\includegraphics[width=1.0\columnwidth]{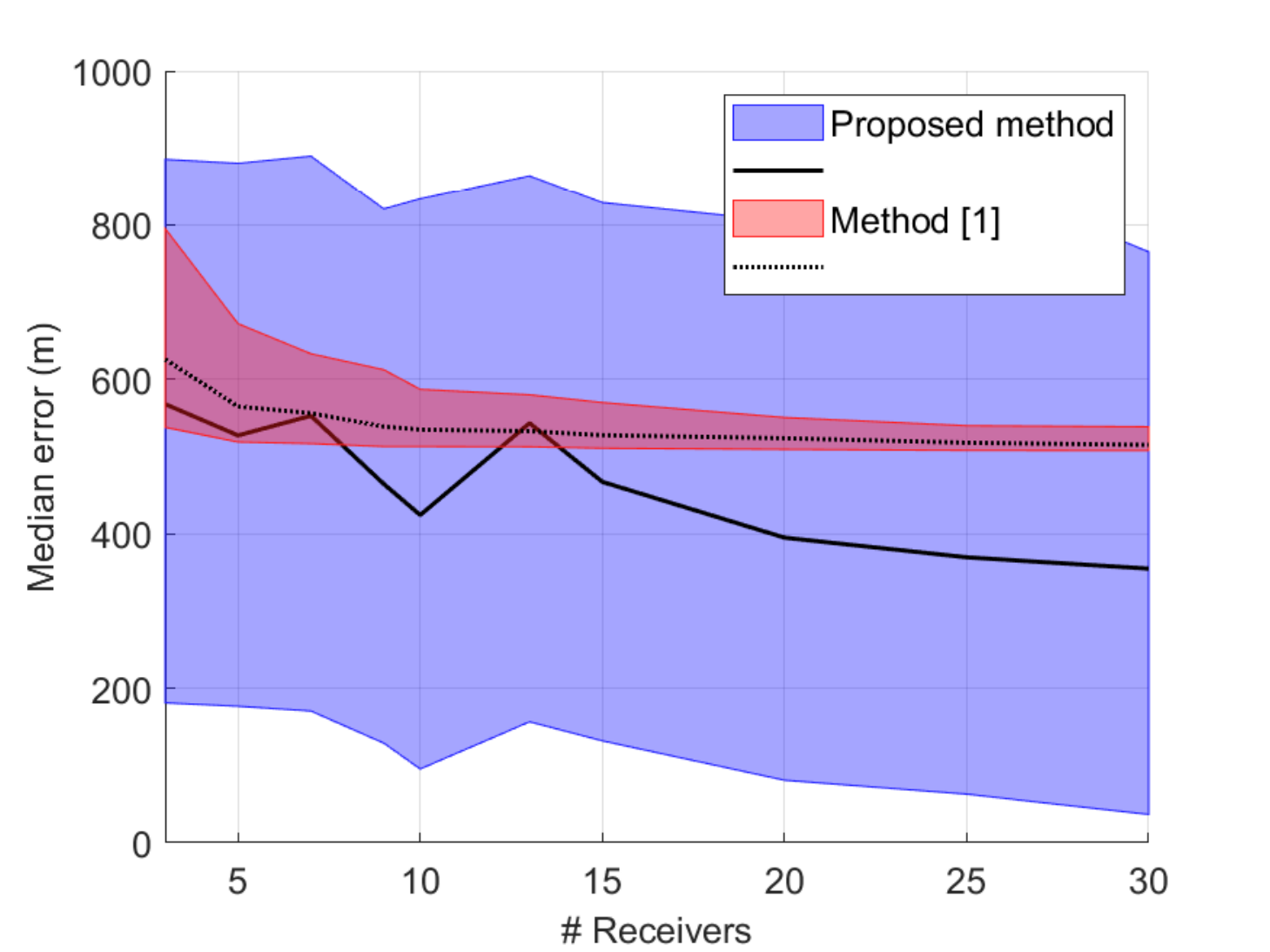}
	\caption{Median 2D-error (black lines) and 25th to 75th percentile (shaded 
	areas) for the proposed method and method \cite{Borio-GNSS+-2016}, using 
	different number of receivers. Receivers travel along a straight 
	line, 500 meters east of the origin.}
	\label{fig:road}
	\vspace{-6pt}
\end{figure}

The performance dependence on the number of receivers is evaluated in this 
section. Ten scenarios, each with a different number of receivers, are tested, 
from 3 to 30 receivers.
Two different mobility scenarios are evaluated, and results are shown in 
Figures~\ref{fig:receivers} and \ref{fig:road} respectively. In the first 
scenario, the receivers' starting positions are placed uniformly inside a cube. 
In the second scenario, all receivers are traveling along a straight line going 
in a north-south direction 500 meters east of the jammer.
The measurement noise variance is set to 
$\sigma_{G,i}^{2} = 0.1$, and the speed of the receivers is 
\SI{1.5}{\meter/\second}. The total time of the scenario is 3000 seconds, 
with a total of $N = 200$ observation samples per receiver. A total of 1000 
Monte Carlo simulations are made.
 
The results can be seen in Figure \ref{fig:receivers}, which shows the median 
estimation error as a function of the number of receivers. It can be seen that 
the proposed algorithm using AGC achieves smaller localization error than the 
state-of-the-art algorithms we compare to. For example, the proposed 
algorithm achieves a median 3D-error of less than 5 meters when using AGC 
mesurements in the simulated scenario with 10 receivers. The corresponding 
errors are 698 meters and 217 meters when using method \cite{Ahmed-ICL-GNSS-2020} 
and \cite{Borio-GNSS+-2016}.

What also can be seen from Figure~\ref{fig:receivers} is that the error 
decreases quickly with the number of receivers. Going from 3 to 5 receivers 
decreases the 75th percentile of the 3D-error from 400 meters to 15 meters 
when using AGC. Such big decreases are not seen in the compared methods, which 
gives our method an advantage in the sense that it would not require as many 
receivers to get a good estimate.

While the error of method \cite{Borio-GNSS+-2016} decreases with increasing number of
 receivers, this might be a bit misleading. In this testcase, the receivers 
are placed using a uniform distribution over the area surrounding the jammer, and with added receivers the centroid 
moves towards the origin, where the jammer is placed. If the jammer is placed 
somewhere else, the result would be worse. This problem is observed in a scenario
where all receivers are traveling along a straight line going in a north-south direction, 
500 meters east of the jammer. All receivers are placed at the same height as the 
transmitter. 

The result is shown in Figure~\ref{fig:road}. The error shown is the 
two-dimensional error, not taking the elevation in consideration, since the receivers 
have the same elevation as the transmitter in this scenario. The error 
never gets smaller than 500 meters for algorithm \cite{Borio-GNSS+-2016}, while the 
proposed method achieves a smaller error in a majority of the simulations, even 
though it is not as accurate as when the receivers are placed uniformly. The loss 
in accuracy comes from the fact that all receivers are located on the same side of the 
jammer, and hence the receiver geometry is worse than if they are spread around the 
jammer.

The algorithm \cite{Ahmed-ICL-GNSS-2020} has a hard time finding 
a solution to the least squares problem during the road scenario, not giving a solution at all for 68\% of 
the 
simulations for 30 receivers and 99\% of the simulations for 3 receivers, and 
is therefore not shown in Figure \ref{fig:road}. 

\subsection{Varying Noise Power}\label{sec:noise}
\begin{figure}[tbp]
	\includegraphics[width=1.0\columnwidth]{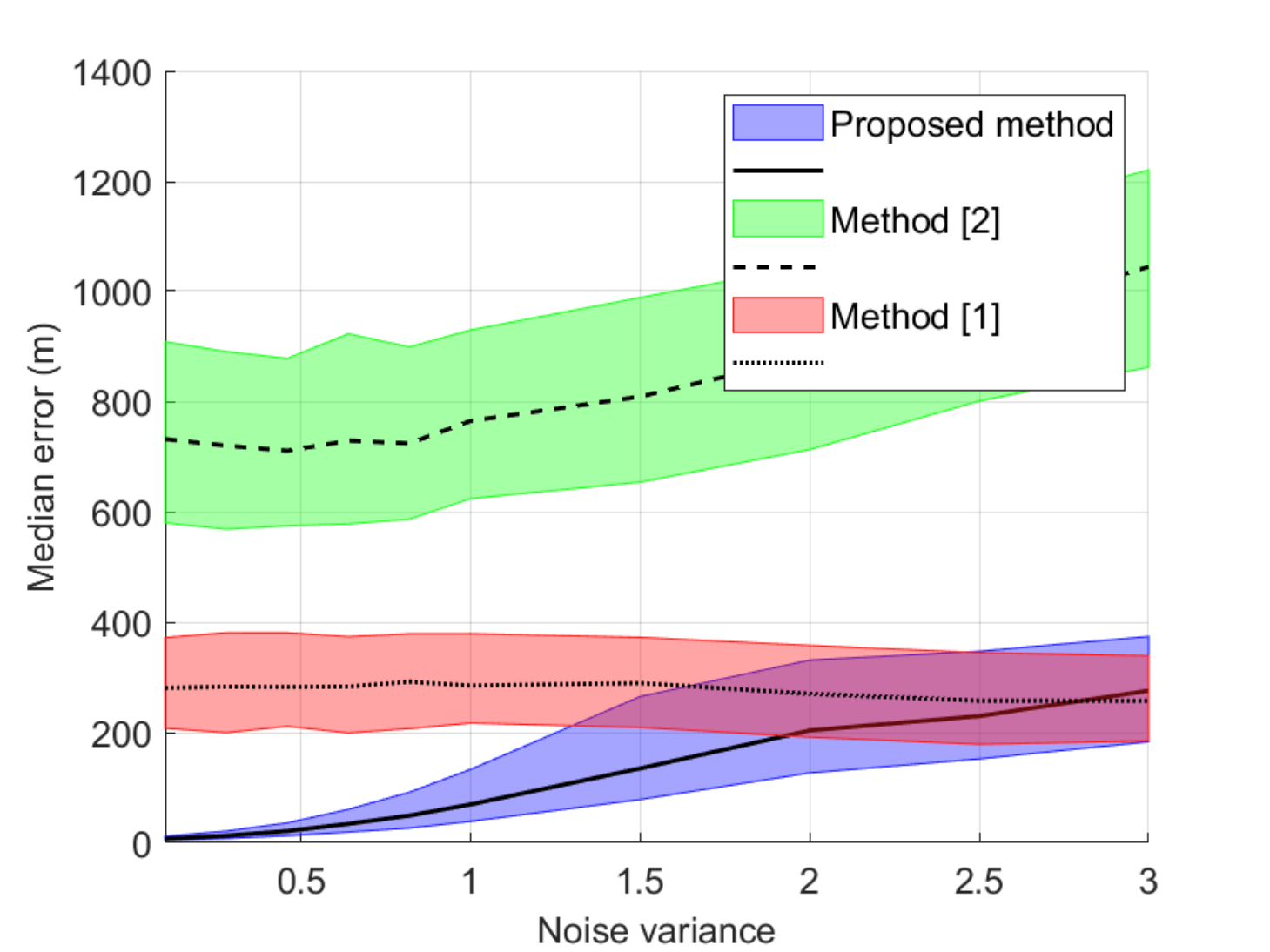}
	\caption{Median 3D-error (black lines) and 25th to 75th percentile (shaded 
	areas) for the three different methods, for different measurement noise 
	powers.}
	\label{fig:noise}
\end{figure}

The localization error dependence of increased measurement noise is evaluated 
in the following. Ten different noise power levels are evaluated, with 
$\sigma_{G,i}^{2}$ ranging from 0.1 to 3, and seven 
receivers are used. Receiver speed and scenario time are not changed from the test in 
Section \ref{sec:receivers}. The results of the simulations can be seen in 
Figure~\ref{fig:noise}. 

Figure~\ref{fig:noise} shows that our proposed algorithm, together with the method 
\cite{Ahmed-ICL-GNSS-2020}, perform worse with increasing noise power while 
method \cite{Borio-GNSS+-2016} does not. Method \cite{Ahmed-ICL-GNSS-2020} 
yields an error that is around 700 meters larger than the error of the proposed 
algorithm throughout the tested noise interval, while the error of the proposed 
algorithm rises to a similar level to that of method \cite{Borio-GNSS+-2016} 
when the noise variance increases. 
The reason that the algorithm of \cite{Borio-GNSS+-2016} is independent of 
measurement noise is that it exploits the receiver positions only, and the 
centroid of the jammed receivers does not depend on the noise power.

\subsection{Initial Value of $\bszeta$} \label{sec:zeta}
\begin{figure}[tbp]
	\includegraphics[width=1.0\columnwidth]{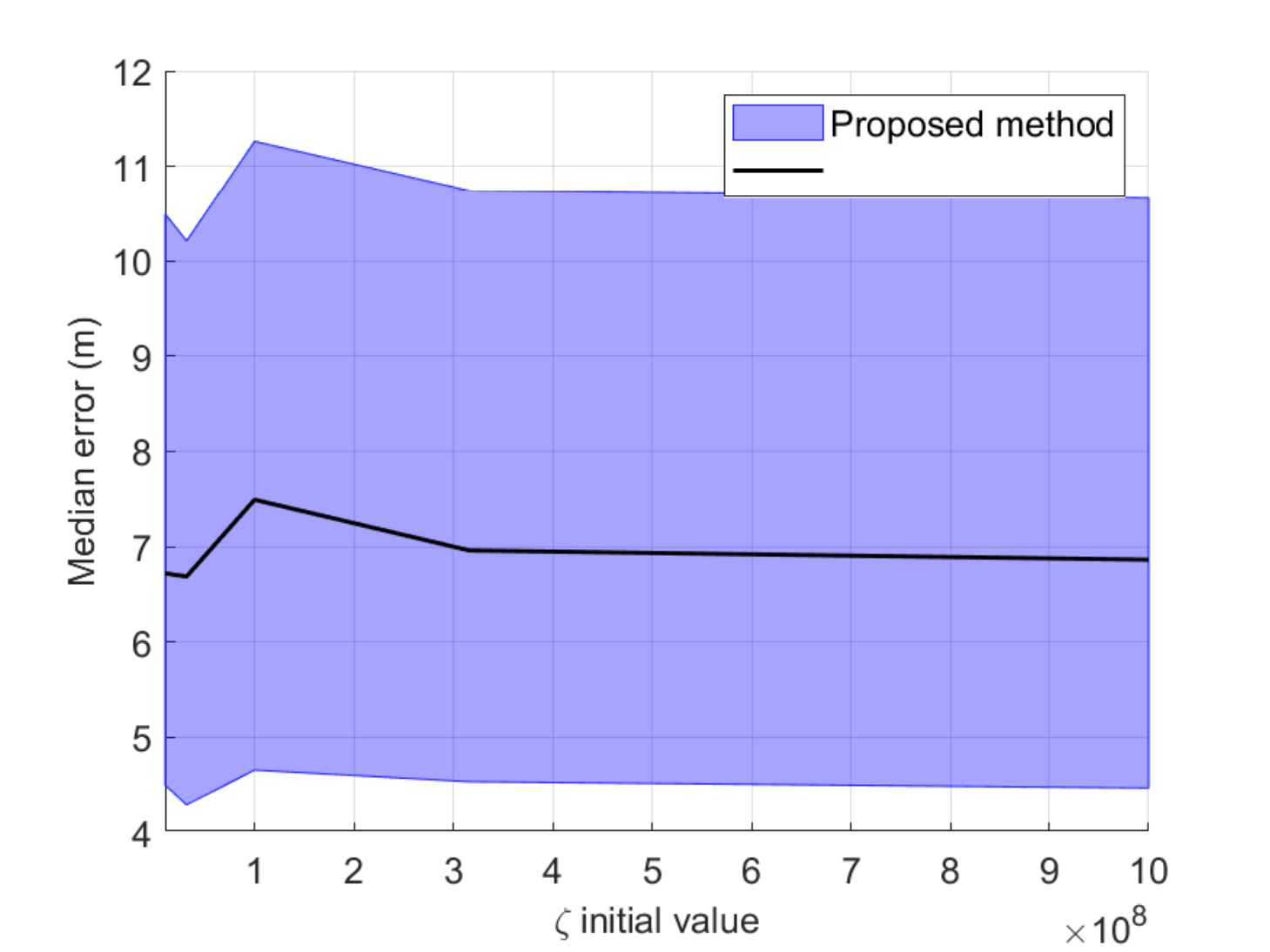}
	\caption{Median 3D-error (black lines) for our method using different initial values on $\zeta$, together with the 25th to 75th percentile (shaded areas).}
	\label{fig:zeta}
\end{figure}

The initial value of the parameter $\bszeta$ in the optimization is difficult 
to set adequatly without further information. To check how much the initial 
value of $\bszeta$ affects the result, a test is made where the gradient descent 
is initiated with different values of $\bszeta$. Five values between \SI{e7}{} 
and \SI{e9}{} are tested. Seven receivers are used, otherwise the settings 
used in Section \ref{sec:receivers} are unchanged. The results can be seen in 
Figure~\ref{fig:zeta}. 

Figure \ref{fig:zeta} shows that the initial value of $\bszeta$ does not affect the 
error much, and can therefore be set rather arbitrarily.

\subsection{CNIR results}
\begin{figure}[tbp]
	\includegraphics[width=1.0\columnwidth]{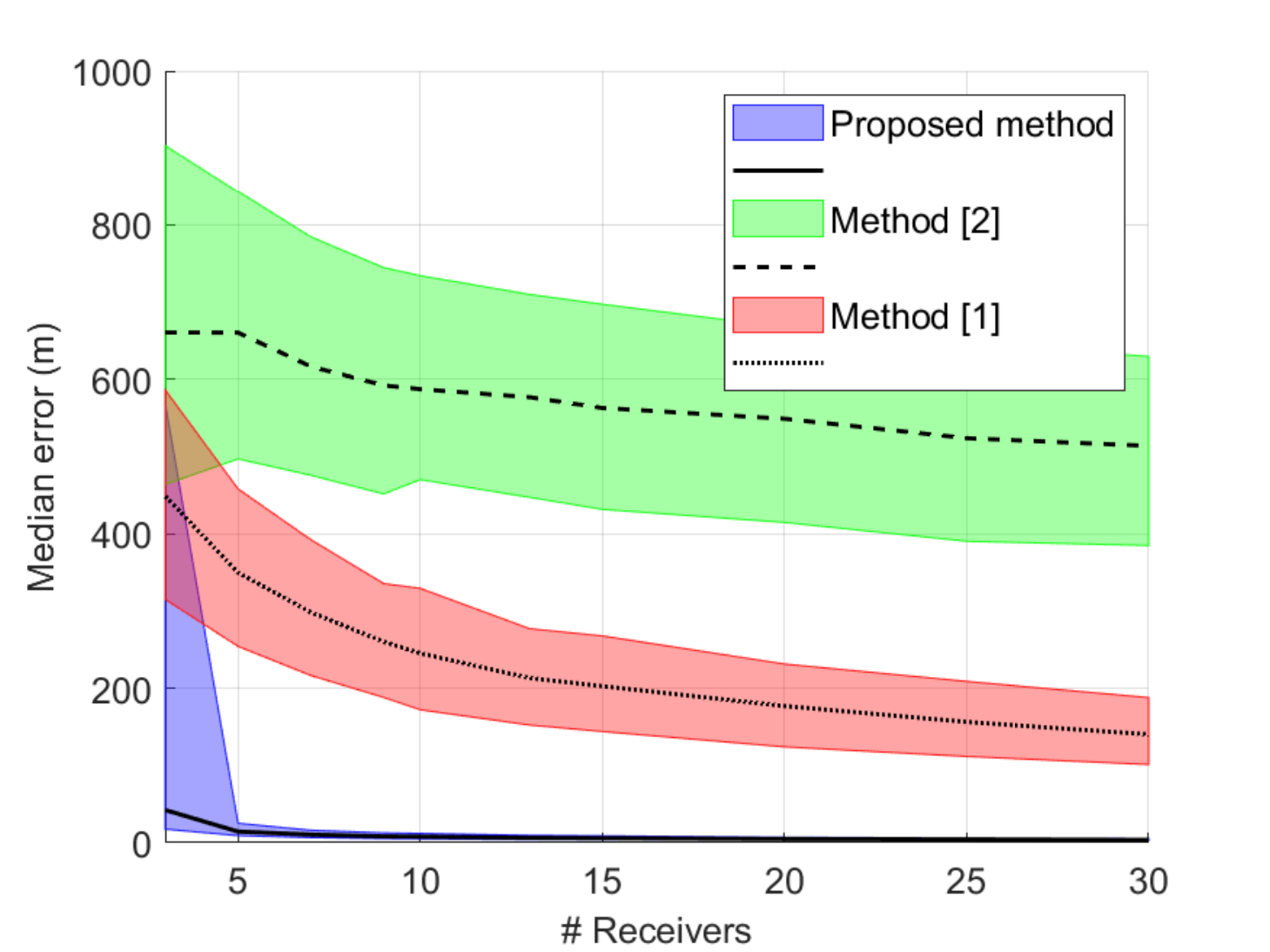}
	\caption{Median 3D-error (black lines) and 25th to 75th percentile (shaded 
	areas) for the three different methods, using different number of 
	receivers.}
	\label{fig:receivers_CNIR}
	\vspace{-6pt}
\end{figure}

\begin{figure}[tbp]
	\includegraphics[width=1.0\columnwidth]{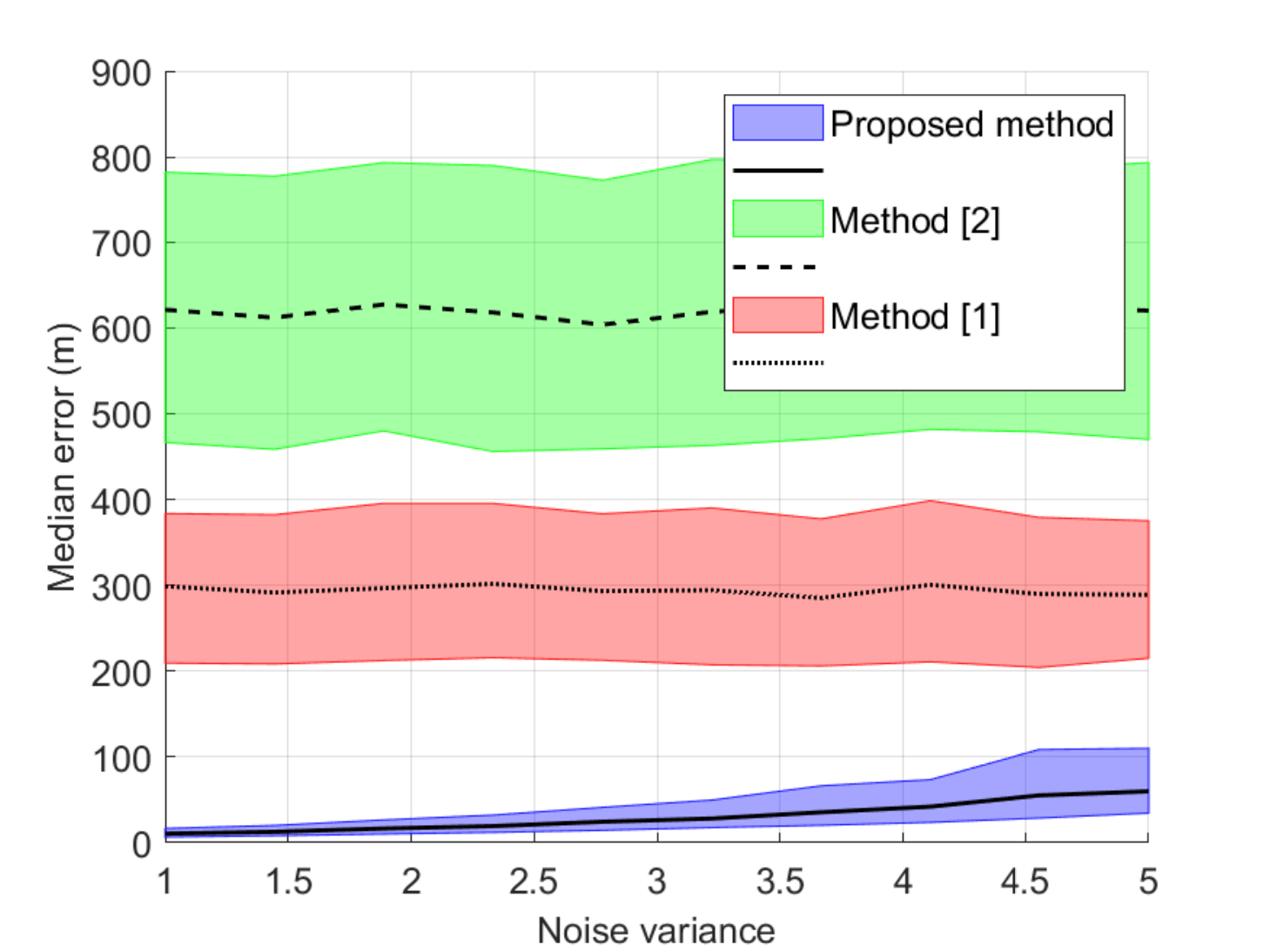}
	\caption{Median 3D-error (black lines) for the three 
	different methods using different noise powers, together with the 25th to 75th percentile (shaded areas).}
	\label{fig:noise_CNIR}
\end{figure}

For the model using CNIR values, the same tests as in Section \ref{sec:receivers} 
(excluding the road scenario) and Section \ref{sec:noise} were executed, testing 
both how the number of receivers and the measurement noise power affect the 
result. The same settings as for the AGC are used, except that the noise power is 
set to $\sigma_{S_{i,j}}^{2} = 1$ and the range when testing different noise power 
levels is changed, going from 1 to 5. The results can be seen in Figure 
\ref{fig:receivers_CNIR} and Figure \ref{fig:noise_CNIR}, which show similar 
results as previously shown for the AGC-based algorithms. The methods are not 
affected by the noise as much as before, because the noise is added on each 
satellite signal, and only the mean of all satellite signals is used.

\section{Concluding Remarks}
\label{sec:conclusion}

The proposed jamming localization algorithm, based on
participatory sensing using AGC and $C/N_0$ estimates from commercial
GNSS receivers, were shown to perform well in relevant scenarios. It was shown 
to outperform similar state-of-the-art localization algorithms, that are also 
based on participatory use of embedded GNSS receivers.
The proposed method gives an estimation error below 15 meters using 
AGC or CNIR measurements from only 5 receivers, compared with hundreds of meters 
for the compared similar state-of-the-art localization methods. This is done without any prior knowledge of the 
jammer or path loss model.

Participatory receivers can be located anywhere and usually not in a 
controllable manner. It was shown that the proposed algorithm performs well for 
different receiver positions, placed randomly around the jammer or along a 
straight line, such as a road, next to the jammer. It should be noted that the proposed model has been 
evaluated with simulated data only, and the next step should be to test it using real measurements from GNSS receivers 
in real life scenarios.


\end{document}